\documentclass[aps, twocolumn, showpacs]{revtex4}

\begin{document}
\title{Comment on 'Single-slit electron diffraction with Aharonov-Bohm phase: Feynman's thought experiment with quantum point contacts'}
\author{S. C. Tiwari \\
Department of Physics, Banaras Hindu University, Varanasi 221005, \\ Institute of Natural Philosophy, 
Varanasi India\\}
\begin{abstract}
We comment on the paper published in PRL 112, 010403 (2014).
\end{abstract}
\maketitle

Aharonov-Bohm (AB) phase has profound role in establishing the physical reality of electromagnetic potentials \cite{1}. One of the important observable consequences is the phase shift due to the enclosed magnetic flux encircled by two coherent electron beams in a typical double-slit interference experiment; the crucial point is that the magnetic flux has to be confined in a small region to ensure field-free passage of the electron beams. Advances in technology have led exploring many variants of the AB effect with increasing precision \cite{2}; the Letter reporting results of a diffraction experiment on ballistic electrons using quantum point contacts \cite{3} would have been a welcome addition in this context but for the conceptual flaws embodied in its presentation.

Authors refer to wave nature of electrons and AB phase as 'two edifices of the quantum theory of matter', however they incorrectly state that 'both of these have been independently tested in the laboratory' since the interference experiments testing AB phase depend on the wave nature of electrons. More serious is their misleading interpretation of the so called Feynman's thought experiment \cite{4}. Recalling that AB phase is a nonlocal effect of potentials where no magnetic field exists significance of three distinct cases of double-slit experiment considered in \cite{4} could be appreciated: schematically illustrated in Figure 15-5 for wave nature of electrons, in Figure 15-7 for the AB effect and in Figure 15-8 for the effect of a weak magnetic field. The last one constitutes the subject matter of \cite{3}. From the outset it must be emphasized that Feynman's motivation to analyse this case is to seek quantum-classical correspondence. A uniform weak magnetic field in a narrow strip extended over a large region behind the slits is imagined and four logical steps comprise the discussion: I) Calculate phase shift due to the flux of the magnetic field behind the wall separating the slits. Since this magnetic field is not directly felt by the electron beams the phase shift is rightly termed the AB phase. II) Assuming the notion of a classical trajectory this phase shift is related to angular deflection on the screen. III) Begin with a classical picture and consider the magnetic field existing adjacent to the slits to calculate the transverse force on the electron beams, and determine the angular deflection. IV) Substitute de Broglie wavelength for the momentum in step III proving the equivalence of the two angular deflections.

Obviously Feynman's arguments are semiclassical. In fact, contrasting them with his treatment of double-slit quantum experiment in \cite{5} it immediately follows that strictly adhering to the quantum picture, the interference phenomenon should disappear once particle trajectories applying magnetic field are identified. The most puzzling aspect that somehow escaped Feynman's attention is the following. The angular deflection in step II has origin in the nonlocal AB effect when the magnetic flux is not intercepted by the electron beams. On the other hand in step III direct interaction with a magnetic field present adjacent to the slits is responsible for the angular deflection. Fortuitously they are formally identical but no inference could be drawn on the quantum to classical correspondence.

In the light of the preceding discussion I offer three comments on the contents of the Letter \cite{3}. (i) The statements like 'He shows that the addition ... for classical particles' in the Abstract, and 'An interplay of these ... weak magnetic field' in the first paragraph are confusing on the real import of Feynman's thought experiment. (ii) Since the authors state that Feynman's observations form the basis of their measurements drastic revision of the physical interpretation of the interesting experiment reported by them is inevitable in view of the puzzling aspect noted above. And (iii) The claimed equivalence of 'the abstract quantum formulation' and 'classical picture' in the concluding part of the Letter is untenable. A thorough reappraisal of Feynman's arguments is necessary; the idea of modular momentum \cite{6} could prove to be useful for this purpose..

\end{document}